\documentclass[runningheads]{llncs}

\usepackage[utf8]{inputenc} 
\usepackage[T1]{fontenc}    
\usepackage{hyperref}       
\usepackage{url}            
\usepackage{booktabs}       
\usepackage{amsfonts}       
\usepackage{nicefrac}       
\usepackage{microtype}      
\usepackage{multirow}
\usepackage{amsmath}
\usepackage[caption=false]{subfig}

\usepackage{graphicx}

\usepackage[dvipsnames]{xcolor}

\usepackage{algorithm,algcompatible,amsmath}
\algnewcommand\INPUT{\item[\textbf{Input:}]}%
\algnewcommand\OUTPUT{\item[\textbf{Output:}]}%

\begin{document}
\title{Deform, Cut and Tear a skinned model using Conformal Geometric Algebra} 

\author{Manos Kamarianakis\inst{1,2}\orcidID{0000-0001-6577-0354}
\and \\ 
George Papagiannakis\inst{1,2}\orcidID{0000-0002-2977-9850}}
\authorrunning{M. Kamarianakis, G. Papagiannakis}
\institute{University of Crete, Greece \and ORamaVR, \url{http://www.oramavr.com} \email{\{m.kamarianakis,george.papagiannakis\}@gmail.com}}

\maketitle

\begin{abstract}
  
  In this work, we present a novel, integrated rigged character simulation 
  framework in Conformal Geometric Algebra (CGA) that supports, for the first time, real-time cuts and tears, before and/or after the animation, while 
  maintaining deformation topology. The purpose of using CGA is to lift 
  several restrictions posed by current state-of-the-art character animation 
  \& deformation methods. Previous implementations originally required 
  weighted matrices to perform deformations, whereas, in the current 
  state-of-the-art, dual-quaternions handle both rotations and translations, 
  but cannot handle dilations. CGA is a suitable extension of 
  dual-quaternion algebra that amends these two major previous shortcomings: 
  the need to constantly transmute between matrices and dual-quaternions as 
  well as the inability to properly dilate a model during animation. Our CGA 
  algorithm also provides easy interpolation and application of all 
  deformations in each intermediate steps, all within the same geometric 
  framework. Furthermore we also present two novel algorithms that enable 
  cutting and tearing of the input rigged, animated model, while the output 
  model can be further re-deformed. These interactive, real-time cut and 
  tear operations can enable a new suite of applications, especially under 
  the scope of a medical surgical simulation.

\end{abstract}



\section{Introduction} 
\label{sec:introduction}

Skinned model animation has become an increasingly important 
research area of Computer Graphics, especially due to the huge 
technological advancements in the field of Virtual Reality and 
computer games. 
The original animation techniques, based on matrices \cite{Alexa:2002ij} for 
translation, rotation and dilation, are still applied as the 
latest GPUs allow for fast  
parallel matrix operations. The fact that the interpolation result 
of two rotation matrices does not result in a rotation matrix, 
forced the use of quaternions as an intermediate step. The extra 
transmutation steps from matrix to quaternions and vice versa, adds
some extra performance burden to the animation but yields 
better results, solving problems such as the gimbal lock.

Nowadays, the state-of-the-art methods for skinned 
model animation use dual-quaternions, an algebraic extension of quaternions \cite{Kenwright:2012tl}. Dual quaternions
handle both rotation and translation, while the dilation effect is 
still applied via matrices \cite{Kavan2008}. It is also noteworthy 
to mention that quaternions and dual quaternions enable 
blending techniques that resolve artifacts produced by simple linear
blending, while further post-processing can be used to further 
minimize them \cite{Kim:2014gb}. 

Advances in Virtual Reality technology and the mass production of 
cheap VR headsets increased the demand of real-time simulation
applications for both personal and industrial purposes. The research areas that sprout from these advancements, 
such as Virtual Surgery Simulation, require more complex model 
deformation such as cutting, tearing or drilling. Current 
algorithms \cite{Bruyns:2002jc,CutSurvey} handle such deformations 
using tetrahedral mesh representation of the model, which demands 
a heavy pre-processing to be performed. Since originally introduced, 
cutting methods have been upgraded and polished to allow 
real-time results, using mostly finite element methods and 
clever optimization \cite{Bielser:gh,Mor:2000jj,Bruyns:2001ki}. 
To make the final results even more realistic, physics engines utilizing 
position-based dynamics are used to simulate soft-tissue cuts 
at the expense of performance 
\cite{Bielser:1999ez,Bender2014,Berndt2017}.

Our approach utilizes the Conformal Geometric Algebra (CGA) 
framework to perform both model animation and cutting. CGA is 
an algebraic extension of dual-quaternions, where all entities such 
as vertices, spheres, planes as well as rotations, translations and dilation 
are uniformly expressed as \emph{multivectors} 
\cite{DietmarFoundations,DorstBook,Wareham:2004js}. The usage of multivectors 
allows model animation without the need to constantly transmute
between matrices and (dual) quaternions, enabling dilation to 
be properly applied with translation \cite{Papagiannakis:2013va,Papaefthymiou:2016dx}. Furthermore, the interpolation of two 
multivectors of the same type correctly produce the expected 
intermediate result \cite{Hadfield:2019cx}, which makes 
creation of keyframes trivial to implement. Finally, usage of the
proposed framework demands a single representation type for all 
data and results, which is the current trend in computer graphics 
\cite{Muller:2016kt}.

\textbf{Our contribution:} The novelty of our work initially involves 
the complete implementation of rigged model animation in terms of CGA, 
extending the work of Papaefthymiou et al. \cite{Papaefthymiou:2016dx} 
with full python-based 
implementation that enables keyframe generation on-the-fly . The 
original animation equation involving matrices is translated to 
its equivalent multivector form (see Section~\ref{sub:multivector_form_of_the_animation_equation}) and all information required to apply the 
formula (vertices, animation data) is obtained from the model and 
translated to multivector. This enables us to have future animation 
models in CGA representation only, which, in combination with an 
optimized GPU multivector implementation, would produce faster results
under a single framework. A novelty of our work 
is the cutting and tearing algorithm that is being applied on 
top of the previous framework; given the input animated model,
we perform real-time cuts and tears on the skin and then further 
re-deform the output model. The subpredicates used in these two 
algorithms utilize the multivector form of their input, so they can 
be implemented in a CGA-only framework. Their design was made in such 
a way that little to no pre-processing of the input model is required 
while allowing a future combination with a physics engine. 
Furthermore, using our method, we can  generate our own keyframes
in real-time  instead of just interpolating between pre-defined ones. 
Our all-in-one cpu python implementation is 
able to process an existing animation model (provided in .dae or 
.fbx format) and translate the existing animation in the desired 
CGA form while further tweaks or deformations are available in 
a simple way to perform. Such an implementation is  optimal 
as far as rapid prototyping, teaching and future connection to 
deep learning is concerned. It also constitutes the base for 
interactive cutting and tearing presented in Section~\ref{sub:cutting_and_tearing}.


\section{State of the art} 
\label{sec:state_of_the_art}

The current state of the art regarding skeletal model animation 
is based on the representation of bones animation via 
transformation matrices and quaternions or dual-quaternions. 
Such an implementation allows for efficient and robust 
interpolation methods between keyframes; linear interpolation 
of the quaternions is done in a naive and easy to perceive way. 
A major drawback of such an implementation is that a dilation
method can not be applied as a scaling matrix always refers to the 
origin and not the parent bone \cite{Papaefthymiou:2016dx}.

To be more precise regarding the mechanics 
of the animation process, in the case of a simple animated model,
every bone $b_i$ amounts to an offset matrix $O_i$ and an 
original transformation matrix $t_i$. The skin of the model is 
imported as a list of vertices $v$ and a list of faces $f$. 
A bone hierarchy is also provided where $\{t_i\}$ are stored along 
with information regarding the animation of each joint. This 
information, usually referred to as \emph{TRS data}, is provided in 
the form of a quaternion, a translation vector and a scaling vector that
represent respectively the rotation, displacement and 
scaling of the joint with respect to the parent joint for each 
keyframe (see Section~\ref{sub:state_of_the_art_representation}).

In order to determine the position of the skin vertices at any 
given time $k$ and therefore render the scene by triangulating 
them using the faces list, we follow the steps described below. 
Initially, a matrix $G$ is evaluated as the inverse of the 
transformation matrix that corresponds to the root node. 
Afterwards, we evaluate the \emph{global transformation matrix} 
for every bone $b_i$ at time $k$ and denote it as $T_{i,k}$. 
To evaluate all 
$T_{i,k}$, we recursively evaluate the matrix product 
$T_{j,k}t_{i,k}$ where $b_j$ is the parent bone of $b_i$, 
given that $T_{r,k}$ is the identity matrix (of size 4), where $b_r$
denotes the root bone. The matrix $t_{i,k}$ is a transformation matrix 
equal to $t_i$ if there is no animation in the model; in this case, our 
implementation allows to generate the keyframes ourselves in real-time.
Otherwise, $t_{i,k}$ is evaluated as

\begin{equation}
t_{i,k} = TR_{i,k} MR_{i,k} S_{i,k}
\end{equation}
where $TR_{i,k}, MR_{i,k},S_{i,k}$ are the interpolated matrices
that correspond to the translation, rotation and scaling of the 
bone $b_i$ at a given time $k$.

After evaluating the matrices $\{T_{i,k}\}$ for all bones $\{b_i\}$, 
we can evaluate the global position of all vertices at  
time $k$, using the \emph{animation equation}:
\begin{equation}\label{eq:animation}
V_k[m] = \displaystyle \sum_{n\in I_m} w_{m,n}G T_{n,k} O_{n}  v[m]
\end{equation}

where
\begin{itemize}
\item $V_k[m]$ denotes the skin vertex of index $m$ (in  
homogeneous coordinates) at the animation time $k$,
\item $I_m$ contains up to four indices of bones that affect the 
vertex $v[m]$,
\item $w_{m,n}$ denotes the ``weight'', i.e., the amount of 
influence of the bone $b_n$ on the vertex $v[m]$,
\item $O_{n}$ denotes the offset matrix corresponding 
to bone $b_n$, with respect to the root bone,
\item $G$ denotes the inverse of the transformation matrix that 
corresponds to the root bone (usually equals the identity matrix) and
\item $T_{n,k}$ denotes the deformation of the bone $b_n$ at animation 
time $k$, with respect to the root bone.
\end{itemize}

\subsection{State-of-the-art representation} 
\label{sub:state_of_the_art_representation}

The modern way to represent the TRS data of a keyframe is 
to use matrices for the translation and dilation data as well as 
quaternions for the rotation data. Let  $\{TR_i,R_i,S_i\}$, 
denote such data at keyframe $i\in\{1,2\}$, where:
\begin{itemize}
  \item $TR_i=\begin{bmatrix}
    1 & 0 & 0 & x_i\\
    0 & 1 & 0 & y_i\\
    0 & 0 & 1 & z_i\\
    0 & 0 & 0 & 1\\
  \end{bmatrix}$ and 
  $S_i=\begin{bmatrix}
    sx_i & 0 & 0 & 0\\
    0 & sy_i & 0 & 0\\
    0 & 0 & sz_i & 0\\
    0 & 0 & 0 & 1\\
  \end{bmatrix}$ represent the translation 
  by $(x_i,y_i,z_i)$ and the scale by $(sx_i,sy_i,sz_i)$ respectively and 
  \item $R_i$ is a quaternion representing 
  the rotation.
\end{itemize} 

Before quaternions, euler andgles and the derived rotation matrices 
were used to represent rotation data. However the usage of such 
matrices induced 
a great problem: a weighted average of such matrices does not correspond 
to a rotation matrix and therefore interpolating between two states would 
require interpolating the euler angles and re-generate the corresponding 
matrix. This in turn would sometimes lead to a gimbal lock or to `candy-wrapper' artifacts such as the ones presented in \cite{Kavan2008}. 

The usage of quaternions allowed for easier interpolation techniques 
while eradicating such problems. Nevertheless, a transformation 
of the interpolated quaternion to corresponding rotation matrix 
was introduced since the GPU currently handles only matrix multiplications 
in a sufficient way for skinning reasons. Therefore, the interpolation between the two keyframes mentioned above follows the following pattern:
\begin{enumerate}
\item the matrices $TR_a = (1-a)TR_1+aTR_2$ and $S_a = (1-a)S_1+aS_2$ 
are evaluated for a given $a\in [0,1]$,
\item the quaternion $R_a = (1-a)R_1+aR_2$ is determined and finally,
\item the rotation matrix $MR_a$ that corresponds to $R_a$ is 
calculated. 
\end{enumerate}

The interpolated data $TR_a, MR_a$ and $S_a$ are then imported to the GPU
in order to determine the  intermediate frame, based on the equation\eqref{eq:animation}.

Using the method proposed in this paper, all data are represented in 
multivector form. A major implication of this change is that the 
interpolation between two states is done in a more clear and uniform way 
as presented in Section~\ref{sec:our_methodology}. 
This also makes the need to constantly transform a quaternion to a 
rotation matrix redundant, although we are now obliged to perform 
multivector additions and multiplications as well as down project points 
from $\mathbb{R}_{4,1}$ to $\mathbb{R}_{3}$ to parse them to the GPU. 
However, since all our data and intermediate results are in the same 
multivector form, we could (ideally) program the GPU
to implement such operations and therefore  greatly improve performance. 


\begin{figure}[bt!]
\centering
\subfloat[]{\includegraphics[width=0.2\textwidth]{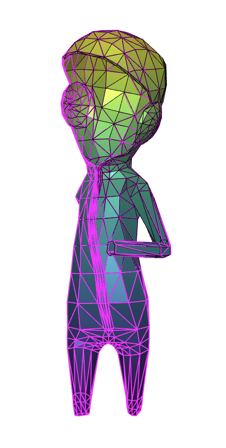}}\hspace*{1cm}
\subfloat[]{\includegraphics[width=0.27\textwidth]{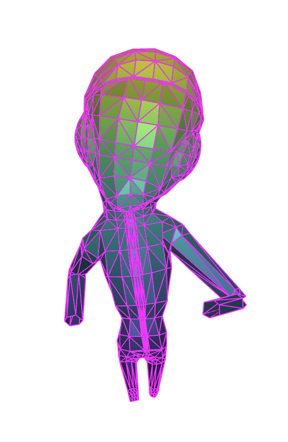}}\hspace*{1cm}
\subfloat[]{\includegraphics[width=0.2\textwidth]{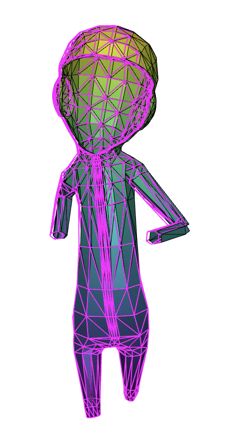}}
\caption{Skinning via multivectors versus skinning via dual quaternions.  The original model is deformed using multivectors and depicted in 
magenta wireframe, superimposed with the color-graded result (based on the $z$ coordinate of each vertex) of the quaternion method for the same deformation. It is qualitatively verified that linear blending of multivectors produces similar results with the current state-of-the-art method. Evaluating the vector differences
of all vertices for the two methods, we have evaluated the approximation error 
assuming the quaternion method to be the correct, using the infinity ($\ell_\infty$) norm. (a) Applying rotation on a bone, approximation error $0.3\%$. (b) 
Applying dilation on a bone, approximation error $0.00035\%$. (c) Applying translation, approximation error $1\%$. The model used contains 1261 vertices and 
1118 faces.}
\label{fig:comparison}
\end{figure}

\section{Our Algorithms and Results} 
\label{sec:our_methodology}


\begin{figure}[bt!]
\centering
\subfloat[]{\includegraphics[width=0.23\textwidth]{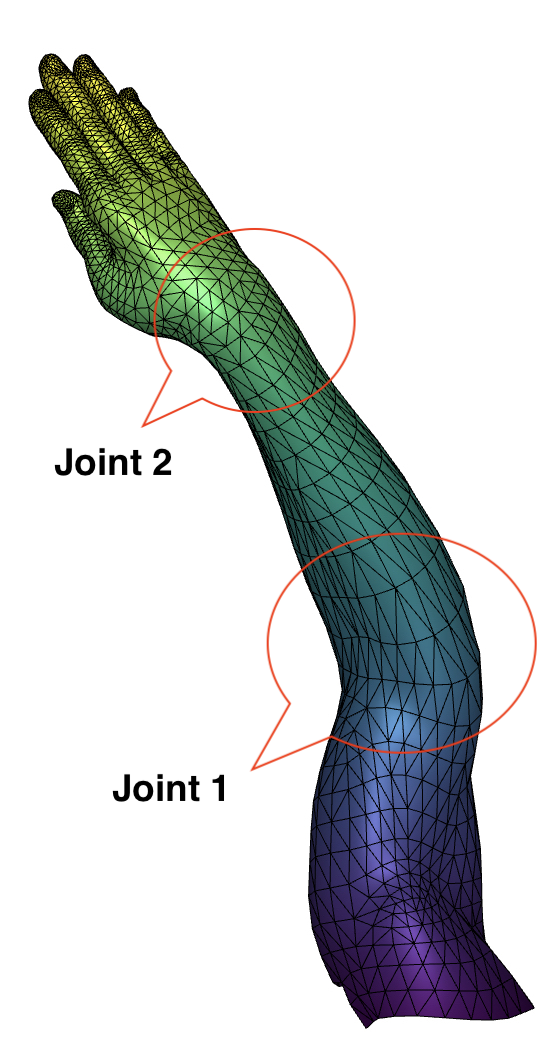}}\hspace*{0mm}
\subfloat[]{\includegraphics[width=0.22\textwidth]{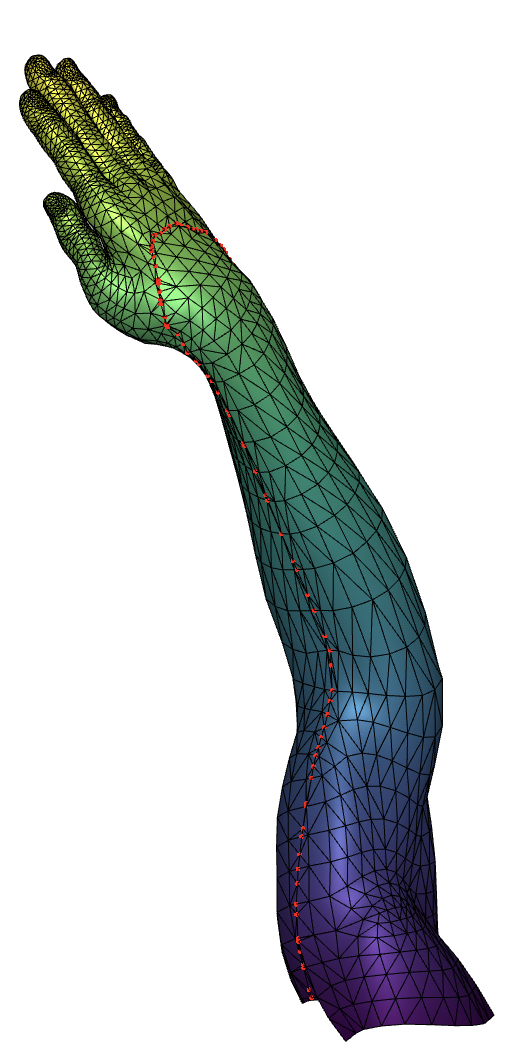}}\hspace*{0mm}
\subfloat[]{\includegraphics[width=0.24\textwidth]{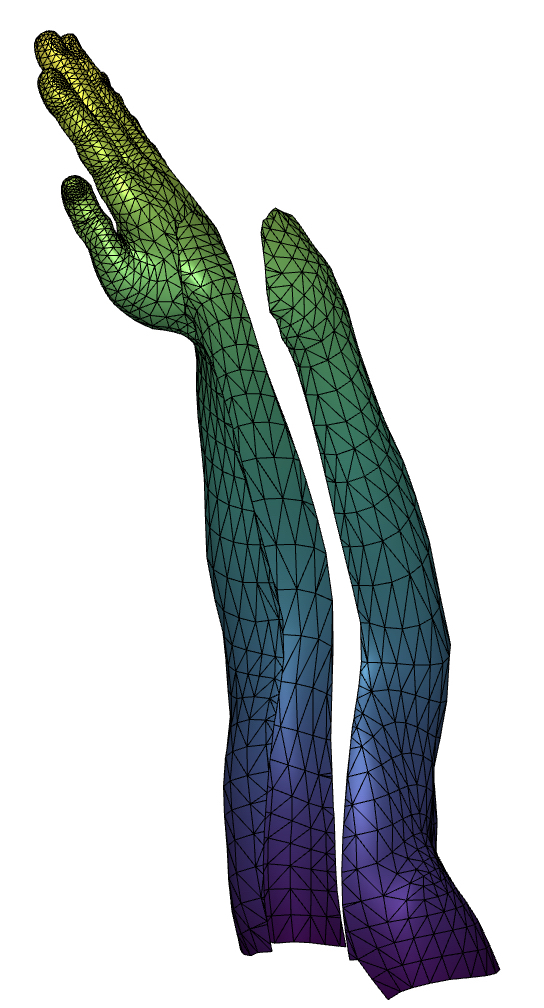}}\hspace*{0mm}
\subfloat[]{\includegraphics[width=0.3\textwidth]{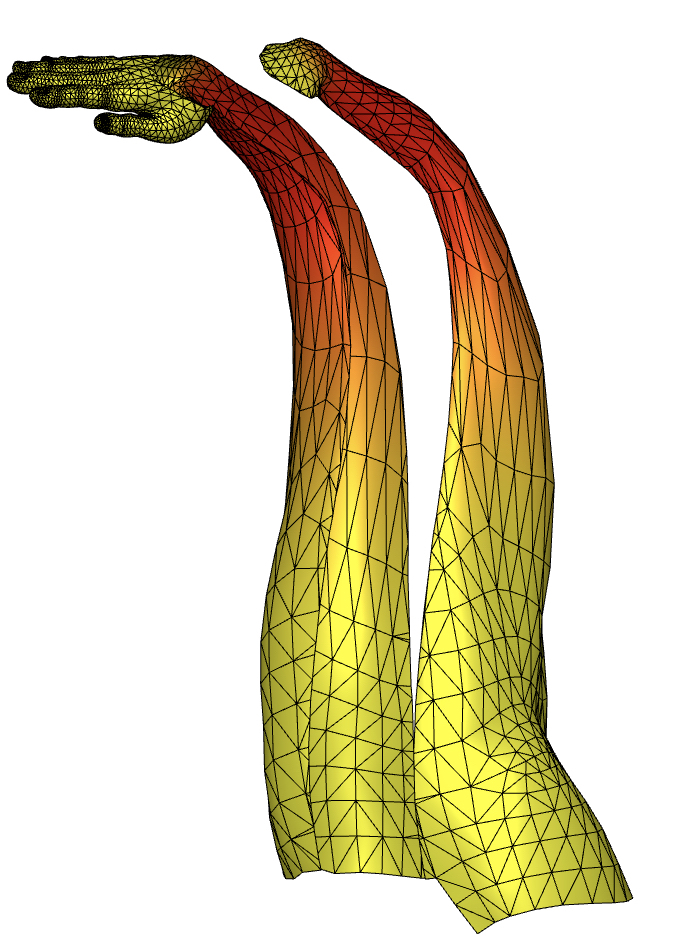}}
\caption{Cutting module intermediate steps. 
(a) The original animated model. (b) The model where the (red) intersection
points of the cutting plane and the mesh are calculated and re-triangulated. (c) The model after the cut. (d) The 
model is deformed by a rotation (axis=$(0,1,1)$, $0.7$ rad), a 
translation (vector=$(13,0,0)$) and a dilation (factor = 0.5)  
at joint 1 (elbow), as well as another rotation (axis=$(0,1,1)$, $0.3$ rad) 
at joint 2 (wrist). Note that 
minimal artifacts occur in the final result. The vertices in (d) are
colored depending on the influence of joint 1 which is mostly deformed. The vertices in (a)-(c) are colored based on their $z$ coordinate.}
\label{fig:cutting_module}
\end{figure}

\subsection{Multivector form of the animation equation} 
\label{sub:multivector_form_of_the_animation_equation}

The animation equation \eqref{eq:animation}, core of the 
animation algorithm, yields fast results (especially when 
combined with a GPU implementation) but denies us a robust way to 
dilate with respect to a bone. Our motivation is to extend and 
apply the animation equation for multivector input as proposed in \cite{Papaefthymiou:2016dx}.

To be more specific regarding our method, we propose the replacement 
of all matrices 
appearing in \eqref{eq:animation} with multivectors for animation 
purposes. The transformation matrix 
of $t_i$ of each bone $b_i$ as well as
all information regarding translation and rotation 
for each keyframe can be easily 
converted to multivectors \cite{DietmarFoundations,DorstBook}. 
Consequently, we can evaluate the multivector $M_{i,k}$ which 
is equivalent to the matrix $T_{i,k}$ by following the same procedure 
of determining the latter (described in 
Section~\ref{sec:state_of_the_art}) while substituting all involved 
matrices with the corresponding multivectors.

Note that various techniques can 
be used to interpolate between two keyframes to obtain $M_{i,k}$; 
for existing keyframes logarithmic blending is preferred 
\cite{Hadfield:2019cx,Kavan2008}, whereas for keyframe generation 
we use linear blending.
In both scenarios, the intermediate results are multivectors of the 
correct type.

Furthermore, each offset matrix $O_n$ and 
each skin vertex $v[m]$ is translated to their CGA form
$B_n$ and $c[m]$ respectively. Finally, $G$ matrix is normalized 
to identity and is omitted in the final equation. 

Our final task is to translate in CGA terms the matrix product 
$T_{n,k} O_n v[m]$, where apparently each 
multiplication sequentially applies a deformation to vertex $v[m]$. 
To apply the respective deformations, encapsulated by $M_{n,k}$ 
and $B_n$, to CGA vertex $c[m]$, we have to evaluate the 
\emph{sandwich geometric product} 
$(M_{n,k}B_{n})c[m](M_{n,k}B_{n})^\star$ where 
$V^\star$ denotes the \emph{inverse} multivector of $V$ (see 
\cite{Kenwright:2012tl,DietmarFoundations} for details).

Summarizing, if the multivector form of the vertex $V_k[m]$, which 
corresponds to the final position of the $m$-th vertex at animation 
time $k$, is denoted by $C_k[m]$, then the \emph{multivector animation 
equation} becomes

\begin{equation}\label{eq:cga_formula}
C_k[m] = \displaystyle 
\sum_{n\in I_m} w_{m,n}(M_{n,k}B_{n})c[m](M_{n,k}B_{n})^\star
\end{equation}
After the evaluation of $C_k[m]$ for all $m$, we can down-project 
all these conformal points to the respective euclidean ones in order to 
represent/visualize them and obtain the final result of the 
keyframe at time $k$.

The replacement of matrices with multivectors enables the introduction 
of dilations in a simple way. The multivector $M_{i,k}$ that represents
a rotation and translation with respect to the parent bone of $b_i$ 
can be replaced with $M_{i,k}D_{i,k}$ where $D_{i,k}$ is the 
corresponding dilator and the operation between them is the geometric
product. The dilator corresponds to a scale factor with respect to the 
parent bone, information that could not be easily interpreted via matrices. 
However, since the application of a motor and/or a dilator to a vertex 
is a sandwich operation, such a dilation becomes possible when 
using multivectors.

A comparison between the results of our proposed method and the 
current state-of-the-art is shown in Figure~\ref{fig:comparison},
where we successfully apply dilation to different bones 
and obtain similar results. Rotations, dilations and translations 
are obtained in our method using multivectors only, 
under a single framework with simpler notation/implementation;
linear blending is used to interpolate between keyframes.


\subsection{Cutting and Tearing Algorithms} 
\label{sub:cutting_and_tearing}


A novelty we present in this paper is the cutting and tearing 
algorithms on skinned triangulated models. As the name suggests, the 
first module enables the user to 
make a planar cut of the model whereas the latter is used to perform 
smaller intersections on the skin. 
In the following sections, we provide a detailed presentation of the 
algorithms involved as well as certain implementation details.

\subsubsection{Cutting Algorithm} 
Cutting a skinned model is implemented in current bibliography in many
forms \cite{Bruyns:2002jc}. The most common technique is via 
the usage of
tetrahedral meshes which require a heavy preprocessing on the model
and  currently do not enable further animation of the model. Our work 
includes an algorithm to planar cut a model (or a part of it) where 
the final mesh is deformable, as we implemented a function to 
calculate weights for all additional vertices that did not originally 
exist (see Figure~\ref{fig:cutting_module}). Most of the subpredicates used in the cutting algorithm are 
implemented in terms of conformal geometry and therefore can be 
used even if the model is provided in multivector form. 

Our proposed planar cut implementation is summarized as 
Algorithm~\ref{alg:cutting}. A description of how we tackle 
the weight evaluation in step~\ref{step:weights_cutting}
is found in Section~\ref{sub:implementation_details}. Our algorithm 
does not require tetrahedral meshed models and requires minimum to none 
preprocessing. It is GA-ready and the low number
of operations it demands make it suitable for VR implementations.

\begin{algorithm}[b]
  \caption{Cutting Algorithm}
  \label{alg:cutting}
  \begin{algorithmic}[1]
    \INPUT Triangulated Mesh $M=(v,f)$ ($f$ is the face list), and a plane $\Pi$. 
    \OUTPUT Two meshes $M_1=(v_1,f_1)$ and $M_2=(v_2,f_2)$, result of $M$ 
    getting cut by $\Pi$  \\
    Evaluate (using GA) and order the intersection points of $\Pi$ with each face of $M$. \\
    Evaluate the weights and bone indices that influence these points.
    \label{step:weights_cutting}\\
    Re-triangulate the faces that are cut using the intersection points.\\
    Separate faces in $f_1$ and $f_2$, depending on which side of the 
    plane they lie. \\
    From $f_1$ and $f_2$, construct $M_1$ and $M_2$. 
  \end{algorithmic}
\end{algorithm}


\subsubsection{Tearing Algorithm} 
  \label{ssub:tearing_module}

\begin{figure}[bt!]
\centering
\subfloat[]{\includegraphics[width=0.2\textwidth]{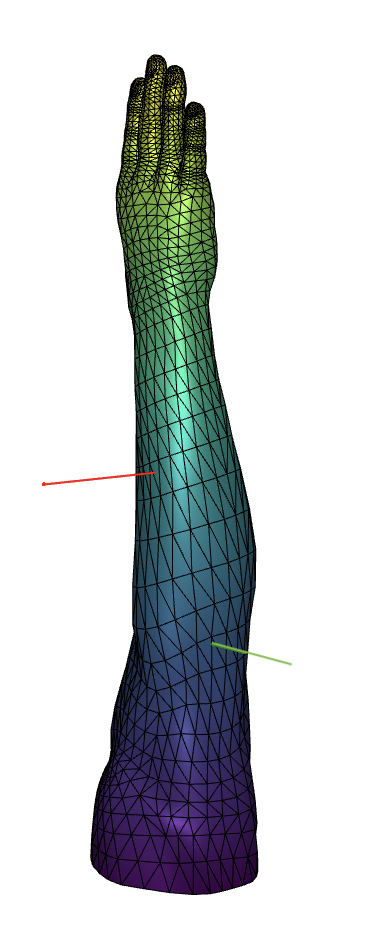}}
\hspace*{10mm}
\subfloat[]{\includegraphics[width=0.2\textwidth]{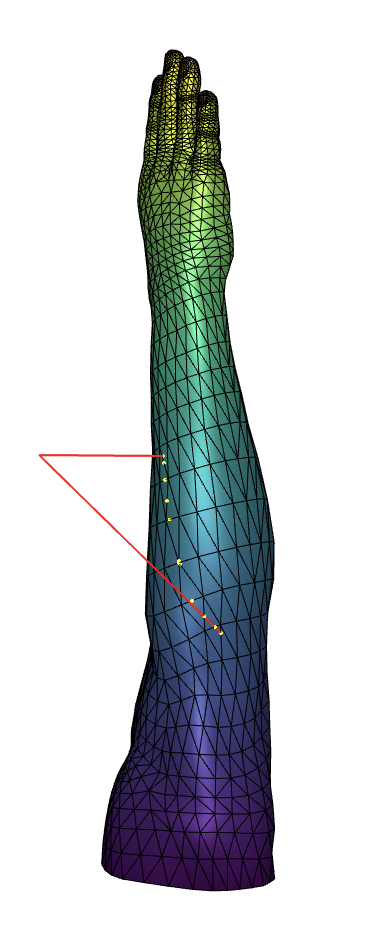}}
\hspace*{15mm}
\subfloat[]{\includegraphics[width=0.2\textwidth]{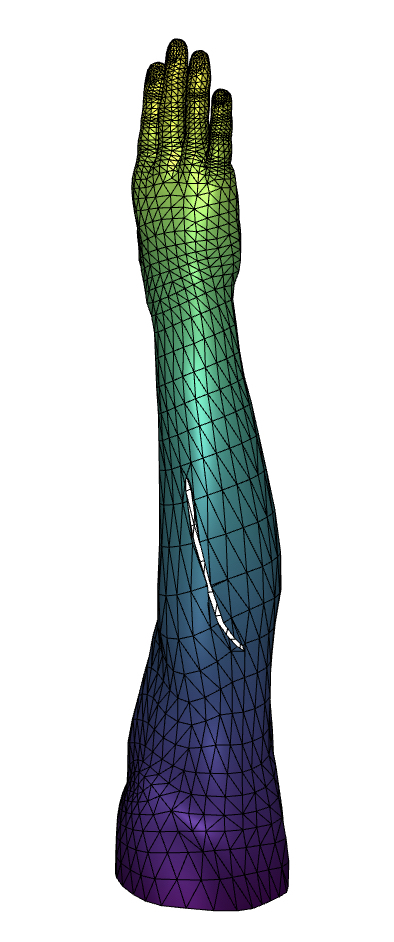}}
\caption{Tearing module intermediate steps. 
(a) The original animated model and the scalpel's position at two 
consecutive time steps.
(b) The plane defined by the scalpels
(depicted as a red tringle) intersects the skin in the magenta points.
(c) The intermediate points are used in the re-triangulation, and 
are «pushed» away from the cutting plane to form an open tear.}
\label{fig:tearing_module}
\end{figure}

\begin{figure}[bt!]
\centering
\subfloat[]{\includegraphics[width=0.15\textwidth]{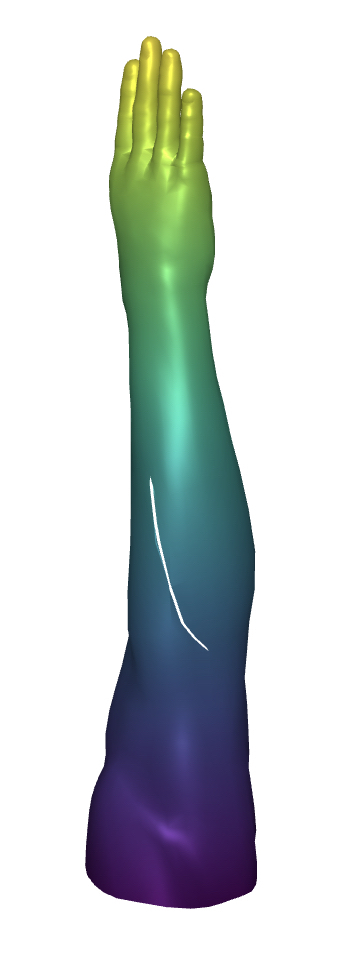}}
\hspace*{10mm}
\subfloat[]{\includegraphics[width=0.11\textwidth]{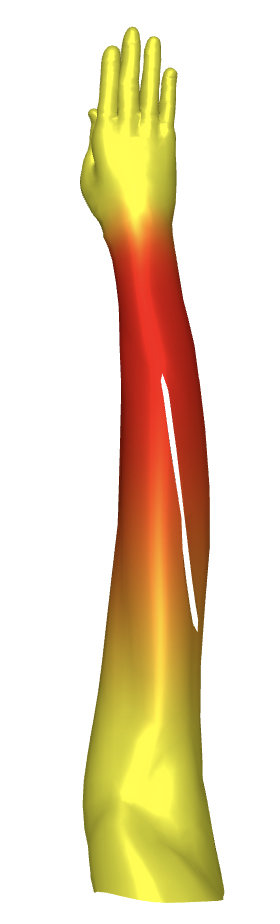}}
\hspace*{10mm}
\subfloat[]{\includegraphics[width=0.2\textwidth]{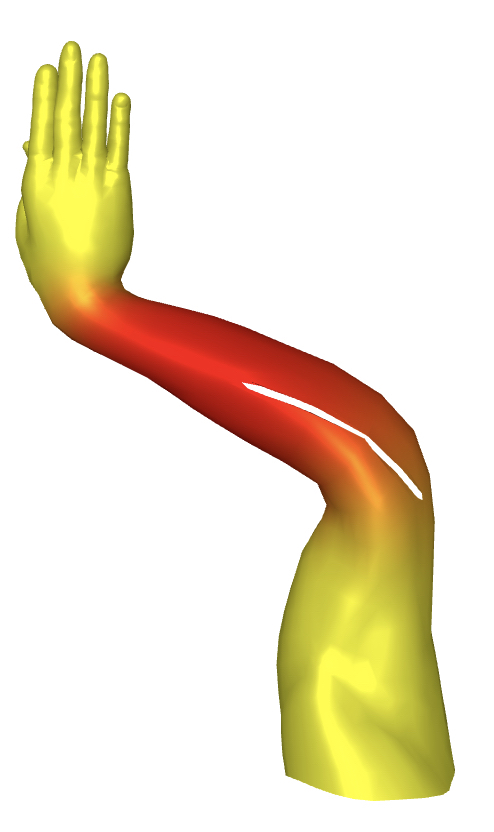}}
\hspace*{10mm}
\subfloat[]{\includegraphics[width=0.12\textwidth]{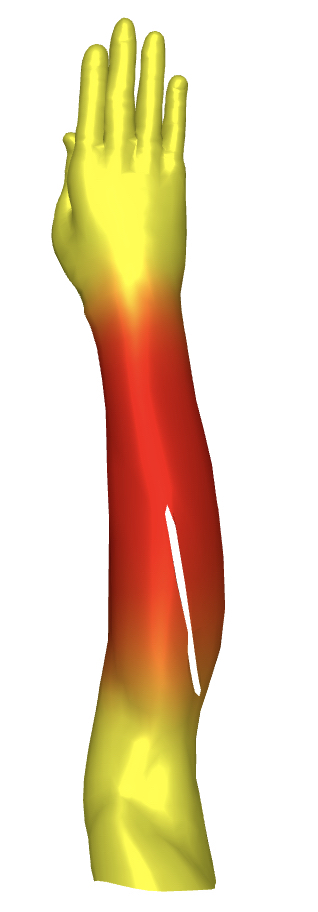}}

\caption{Deformation of a torn model. 
(a) The original model after applying the tear.
(b) Two rotations are applied to the torn model, 
one at elbow joint around $y$-axis by -1 rad, and another at
wrist joint around $y$-axis by 1 rad.
(c) A dilation of scale 1.5 is applied to the torn model,
at elbow joint.
(d) A translation is applied to the torn model at elbow joint with 
translation vector $(18,0,0)$. In all cases, 
minor artifacts only arise, despite the great magnitude of the 
applied deformations. In (b),(c) and (d), vertices are colored 
depending on the influence of elbow joint which is mostly deformed. In (a), vertices are colored based on their $z$ coordinate.}
\label{fig:tearing_module_arm}
\end{figure}

  The purpose of this module is to enable partial cuts on the skinned
  model, in contrast with the cutting module where the cut is, in 
  a sense, complete. The importance of this module derives from the
  fact that most of the surgical incisions are partial cuts and 
  therefore they are worth replicating in the context of a virtual 
  surgery. Towards that direction, our work involves an algorithm 
  that both tears a skinned model and also enables animation of the
  final mesh (see Figures~\ref{fig:tearing_module} and \ref{fig:tearing_module_arm}). 

  To understand the philosophy behind the design of the tearing 
  algorithm that is described below, one must comprehend the 
  differences between cutting and tearing. In tearing, the movement 
  of a scalpel defines the tear rather than a single plane. 
  To capture such a tear in geometric terms, we have to take into
  consideration the location of the scalpel in either a continuous way
  (e.g. record the trail of both endpoints of the scalpel in terms 
  of time) or a discrete way (e.g. know the position of the scalpel 
  at certain times $t_i$). For VR purposes, the latter way is 
  preferred as it yields results with better fps, since input is
  hard to be monitored and logged continuously in a naive way. 
  For these reasons, our implementation requires the scalpel position 
  to be known for certain $t_i$.

  The proposed tearing algorithm is summarized in 
  Algorithm~\ref{alg:tearing}. A description of how we tackle 
  the weight evaluation in step~\ref{step:weights_tearing}
  is found in Section~\ref{sub:implementation_details}.

  \begin{algorithm}
  \caption{Tearing Algorithm}
  \label{alg:tearing}
  \begin{algorithmic}[1]
    \INPUT Triangulated Mesh $M=(v,f)$, and scalpel position 
    at time steps $t_i$ and $t_{i+1}$
    \REQUIRE Scalpel properly intersects $M$ at these time steps
    \OUTPUT The mesh $M_t=(v_t,f_t)$ resulting from $M$ getting torn
    by the scalpel\\
    Determine the intersection points $S_i$ and 
    $S_{i+1}$ of $M$ with the scalpel at time step $t_i$ 
    and $t_{i+1}$ respectively. \\
    Determine the plane $\Pi$, containing 
    $S_i$ and the endpoints of scalpel at time $t_{i+1}$. Small time steps guarantee that $\Pi$ is well-defined. \\
    Evaluate the intersection points $Q_j$ of $\Pi$ and $M$, s.t. 
    the points $S_i$,$Q_0$,$Q_1$,$\ldots$,$Q_m$,
    $S_{i+1}$ appear in this order on $\Pi$ when traversing the 
    skin from $S_i$ to $S_{i+1}$.\\
    Assign weights to points $S_i$, $S_{i+1}$ and all $Q_j$. \label{step:weights_tearing}\\
    Re-triangulate the torn mesh, duplicating $Q_j$ vertices.\\
    Move the two copies of $Q_j$ away from each other to 
    create a visible tear (optional).\label{step:open_tearing}
  \end{algorithmic}
  \end{algorithm}

  Our major assumption is that all intermediate intersection 
  points lie on this plane, which is equivalent to the assume 
  that the tearing curve is smooth, given that $t_i$ and $t_{i+1}$ 
  are close enough. In our implementation, during 
  step~\ref{step:open_tearing}, the intermediate torn points are  
  moved parallel to the direction of the normal of the plane 
  $\Pi$ and away from it, to replicate the opening of a cut 
  human tissue.


\subsection{Implementation Details, Performance and Video Results} 
\label{sub:implementation_details}

The main framework used for skinning and animation with the use of 
multivectors is Python's PyAssimp\footnote[1]{{PyAssimp} Homepage: \url{https://pypi.org/project/pyassimp/}} and Clifford\footnote[2]{{Clifford} Homepage: \url{https://clifford.readthedocs.io/}} package 
for the 
evaluation of the vertices and the Meshplot package for rendering the 
model. The 
use of Python language was preferred for a more user and 
presentation-friendly experience; for a more robust and efficient 
implementation C++ would be advised.

An instance of a class called v\_w is used to store 
for each vertex a list of up to 4 bones that influence it along with 
the corresponding influence factors. The node tree is then traversed 
and all information regarding rotation, translation 
and dilation are translated to multivectors \cite{DietmarFoundations,DorstBook} and also stored in the instance for convenience. 
In order to evaluate the final position of the vertices, all that
is left is to to evaluate the sum in equation~\eqref{eq:cga_formula} for all 
vertices and down project it to $\mathbb{R}^3$, for each vertex. 
There are two possible ways of achieving this task. The first way is 
to evaluate the sum and then 
down project the final result to obtain each vertex in Euclidean form. 
The second way is to down project each term and then add them to get
the final result. Although not obvious, the second method yields 
faster results since the addition of 4 multivectors (32-dimensional arrays)
and one down-projection is slower than down-projecting (up to) 4 
multivectors and adding 4 euclidean vectors of dimension 3.

A final implementation detail regards the weight evaluation for 
newly added vertices in the cutting and tearing modules. In the
former module, such vertices necessarily lie on an edge of the 
original mesh, whose endpoints both lie on different sides of the 
cutting plane. 
Another method is the one used in the tearing 
module where the intersection point can also lie inside a face. 
Assuming the point $X$ lie somewhere on the face $ABC$, we can 
explicitly write $OX = p OA + q OB + r OC$ for 
some $a,b,c\in [0,1 ]$ such that $p+q+r=1$. The tuple $(p,q,r)$ 
is called the \emph{barycentric coordinate} of $X$ with respect to the 
triangle $ABC$. Each of the vertices $A,B,C$ are (usually) 
influenced by up to 4 bones, so let us consider that they are all 
influenced by a set of $N(\leq 12)$ vertices, where the bones 
beside the original 4 have weight 0. Let $w_A,w_B,w_C,w_X$ denote the 
vectors containing the $N$ weights that correspond to 
vertices $A,B,C$ and $X$ respectively, for the same ordering of the 
$N$ involved bones. To determine $w_X$, we first evaluate 
$w = p w_A + q w_B + r w_C$ and consider two cases. 
If $w$ contains up to 4 non-zero weights, then $w_X = w$. Otherwise, 
since each vertex can be influenced by up to 4 bones, we keep the 
4 greater values of $w$, set the others to zero, and normalize the 
vector so that the sum of the 4 values add to 1; the final result
is returned as $w_X$. We denote this weight as \emph{weight of $X$ 
via barycentric coordinates}. Variations of this technique can be 
applied in both modules to prioritize or neglect influences on 
vertices lying on a specific side of the cutting plane. Different 
variations of the weight function allows for less artifacts, 
depending on the model and the deformation subsequent to the 
cutting/tearing.

\textbf{Performance:} Running the Tearing algorithm in the arm model 
(5037 faces, 3069 vertices) it took 2437ms for the final output, 
for 34 intersection points. 
Most of this time (2411ms) were needed just to determine which two faces 
were intersected by the scalpel. Tearing the cylinders model (758 faces,
634 vertices) took 362ms for 17 intersection points. Again, most time
(331ms) was spend for the scalpel intersection. For the Cutting Algorithm,
it took for the cylinders model a total of 898ms:  42ms for vertex separation,
757ms for re-triangulation of the 92 intersection points, 87ms to split 
faces in two meshes and 12ms to update the weights. To cut the arm model, 
it took 22805ms, where most of them (22547ms) were spent to 
re-triangulate the 90 intersection points. These running times
can be greatly improved as our current unoptimized  CPU-based Python 
implementation has to thoroughly search all faces for cuts/tears. A GPU implementation optimized for multivector operations would allows to the comparison of our proposed method with the current state-of-the-art methods.

\textbf{Video:} A video with our results can be found at \url{https://bit.ly/3fsYkdZ}


\section{Conclusions and Future Work} 
\label{sec:conclusions_and_future_work}

This work describes a way to perform model animation and deformation 
as well as cutting and tearing under a single geometric framework called
Conformal Geometric Algebra. Our results were obtained using python but
since our goal is to have a full implementation in real-time virtual reality 
simulation we will inevitably have to embed in C++ and 
ultimately Unity/Unreal Engine code. We intend to combine the tearing 
module in conjunction with a physics engine to obtain a realistic opening 
effect. A drilling module is in progress that will allow 
the user to make holes on the skinned model; such a task is 
useful especially for VR simulations of dental surgeries. 
Finally, it is our intention to minimize running times to 
real-time implementation levels via optimization and the 
use of recently developed acceleration techniques \cite{Hadfield:2019fm}. 


\bibliographystyle{unsrt} 

\bibliography{references}

\begin{thebibliography}{10}

\bibitem{Alexa:2002ij}
Marc Alexa.
\newblock {Linear combination of transformations.}
\newblock {\em ACM Trans. Graph.}, 21(3):380--387, 2002.

\bibitem{Kenwright:2012tl}
Ben Kenwright.
\newblock {A beginners guide to dual-quaternions: What they are, how they work,
  and how to use them for 3D character hierarchies}.
\newblock In {\em WSCG 2012 - Conference Proceedings}, pages 1--10. Newcastle
  University, United Kingdom, December 2012.

\bibitem{Kavan2008}
Ladislav Kavan, Steven Collins, Ji{\v{r}}{\'\i} {\v Z}{\'a}ra, and Carol
  O'Sullivan.
\newblock {Geometric skinning with approximate dual quaternion blending}.
\newblock {\em dl.acm.org}, 27(4), October 2008.

\bibitem{Kim:2014gb}
Young~Beom Kim and Jung~Hyun Han.
\newblock {Bulging-free dual quaternion skinning}.
\newblock In {\em Computer Animation and Virtual Worlds}, pages 321--329. Korea
  University, Seoul, South Korea, John Wiley {\&} Sons, Ltd, January 2014.

\bibitem{Bruyns:2002jc}
C~D Bruyns, S~Senger, A~Menon, K~Montgomery, S~Wildermuth, and R~Boyle.
\newblock {A survey of interactive mesh-cutting techniques and a new method for
  implementing generalized interactive mesh cutting using virtual
  tools{\ddag}}.
\newblock {\em The Journal of Visualization and Computer Animation},
  13(1):21--42, February 2002.

\bibitem{CutSurvey}
Jun Wu, R{\"u}diger Westermann, and Christian Dick.
\newblock {A Survey of Physically Based Simulation of Cuts in Deformable
  Bodies}.
\newblock {\em Computer Graphics Forum}, 34(6):161--187, September 2015.

\bibitem{Bielser:gh}
D~Bielser, P~Glardon, M~Teschner, and M~Gross.
\newblock {A state machine for real-time cutting of tetrahedral meshes}.
\newblock In {\em 11th Pacific Conference on Computer Graphics and
  Applications}, pages 377--386. IEEE Comput. Soc, 2004.

\bibitem{Mor:2000jj}
Andrew~B Mor and Takeo Kanade.
\newblock {Modifying Soft Tissue Models: Progressive Cutting with Minimal New
  Element Creation}.
\newblock In {\em Advances in Computer Graphics}, pages 598--607. Springer
  Berlin Heidelberg, Berlin, Heidelberg, 2000.

\bibitem{Bruyns:2001ki}
Cynthia~D Bruyns and Steven Senger.
\newblock {Interactive cutting of 3D surface meshes}.
\newblock {\em Computers {\&} Graphics}, 25(4):635--642, August 2001.

\bibitem{Bielser:1999ez}
Daniel Bielser, Volker~A Maiwald, and Markus~H Gross.
\newblock {Interactive Cuts through 3-Dimensional Soft Tissue}.
\newblock {\em Computer Graphics Forum}, 18(3):31--38, 1999.

\bibitem{Bender2014}
Jan Bender, Matthias M{\"u}ller, Miguel~A Otaduy, Matthias Teschner, and Miles
  Macklin.
\newblock {A survey on position-based simulation methods in computer graphics}.
\newblock {\em Computer Graphics Forum}, 33(6):228--251, September 2014.

\bibitem{Berndt2017}
Iago~U Berndt, Rafael~P Torchelsen, and Anderson Maciel.
\newblock {Efficient Surgical Cutting with Position-Based Dynamics.}
\newblock {\em IEEE Computer Graphics and Applications}, 37(3):24--31, 2017.

\bibitem{DietmarFoundations}
D~Hildenbrand.
\newblock {\em {Foundations of geometric algebra computing, 2013}}.
\newblock Springer.

\bibitem{DorstBook}
Leo Dorst, Daniel Fontijne, and Stephen Mann.
\newblock {\em {Geometric algebra for computer science - an object-oriented
  approach to geometry.}}
\newblock The Morgan Kaufmann series in computer graphics, 2007.

\bibitem{Wareham:2004js}
Rich Wareham, Jonathan Cameron, and Joan Lasenby.
\newblock {Applications of Conformal Geometric Algebra in Computer Vision and
  Graphics.}
\newblock {\em IWMM/GIAE}, 3519(1):329--349, 2004.

\bibitem{Papagiannakis:2013va}
George Papagiannakis.
\newblock {Geometric algebra rotors for skinned character animation blending}.
\newblock In {\em SIGGRAPH Asia 2013 Technical Briefs, SA 2013}, December 2013.

\bibitem{Papaefthymiou:2016dx}
Margarita Papaefthymiou, Dietmar Hildenbrand, and George Papagiannakis.
\newblock {An inclusive Conformal Geometric Algebra GPU animation interpolation
  and deformation algorithm}.
\newblock {\em The Visual Computer}, 32(6-8):751--759, June 2016.

\bibitem{Hadfield:2019cx}
H~Hadfield and J~Lasenby.
\newblock {Direct Linear Interpolation of Geometric Objects in Conformal
  Geometric Algebra}.
\newblock {\em Advances in Applied Clifford Algebras}, 2019.

\bibitem{Muller:2016kt}
Matthias M{\"u}ller, Nuttapong Chentanez, and Miles Macklin.
\newblock {Simulating visual geometry}.
\newblock In {\em Proceedings - Motion in Games 2016: 9th International
  Conference on Motion in Games, MIG 2016}, pages 31--38, 2016.

\bibitem{Hadfield:2019fm}
Hugo Hadfield, Dietmar Hildenbrand, and Alex Arsenovic.
\newblock {Gajit: Symbolic Optimisation and JIT Compilation of Geometric
  Algebra in Python with GAALOP and Numba}.
\newblock In {\em Advances in Computer Graphics}, pages 499--510. Springer,
  2019.

\end{thebibliography}

\end{document}